\documentclass[12pt,preprint]{aastex}

\newcommand{\teff}{$T_{\rm eff}$} 
\begin{document}

\title{Rubidium and lead abundances in giant stars of the globular 
clusters M 13 and NGC 6752\footnote{Based in part on data 
collected at the Subaru Telescope, 
which is operated by the National Astronomical Observatory of Japan and on 
observations made with the Magellan Clay Telescope at Las Campanas 
Observatory.}}

\author{David Yong}
\affil{Department of Physics \& Astronomy, University of North
Carolina, Chapel Hill, NC 27599-3255}
\email{yong@physics.unc.edu}

\author{Wako Aoki}
\affil{National Astronomical Observatory, Mitaka, 181-8588 Tokyo, Japan}
\email{aoki.wako@nao.ac.jp}

\author{David L.\ Lambert}
\affil{Department of Astronomy, University of Texas, Austin, TX 78712}
\email{dll@astro.as.utexas.edu}

\author{Diane B.\ Paulson}
\affil{NASA's Goddard Space Flight Center, Code 693.0, Greenbelt MD
20771}
\email{diane.b.paulson@gsfc.nasa.gov}

\begin{abstract}
We present measurements of the neutron-capture elements 
Rb and Pb in five giant stars of the 
globular cluster NGC 6752 and Pb measurements in four giants of the 
globular cluster M 13. The abundances were derived by comparing synthetic 
spectra with high resolution, high signal-to-noise ratio spectra 
obtained using HDS on the Subaru telescope and MIKE on the Magellan telescope. 
The program stars span the range of the O-Al abundance variation. In NGC 6752, 
the mean abundances are [Rb/Fe] = $-$0.17 $\pm$ 0.06 ($\sigma$ = 0.14), 
[Rb/Zr] = $-$0.12 $\pm$ 0.06 ($\sigma$ = 0.13), and 
[Pb/Fe] = $-$0.17 $\pm$ 0.04 ($\sigma$ = 0.08). In M 13 the mean abundance 
is [Pb/Fe] = $-$0.28 $\pm$ 0.03 ($\sigma$ = 0.06). Within the measurement 
uncertainties, we find no evidence for a star-to-star variation for either 
Rb or Pb within these clusters. None of the abundance ratios [Rb/Fe], [Rb/Zr], 
or [Pb/Fe] are correlated with the Al abundance. NGC 6752 may have slightly 
lower abundances of [Rb/Fe] and [Rb/Zr] compared to the small sample of field 
stars at the same metallicity. For M 13 and NGC 6752 
the Pb abundances are 
in accord with predictions from a Galactic chemical evolution model.
If metal-poor intermediate-mass asymptotic giant branch stars did produce
the globular cluster abundance anomalies, then such stars do not synthesize
significant quantities of Rb or Pb. Alternatively, if such stars do synthesize
large amounts of Rb or Pb, then they are not responsible for the abundance 
anomalies seen in globular clusters. 
\end{abstract}


\section{Introduction}
\label{sec:intro}

Globular clusters are ideal laboratories for testing the predictions 
of stellar evolution theory (e.g., \citealt{renzini88}) 
since the individual stars are believed to be monometallic, coeval, and at 
the same distance. In a given globular cluster (excluding $\omega$ Cen), 
spectroscopic observations of individual stars have confirmed that members 
have uniform compositions, at least for the Fe-peak elements 
(e.g., see review by \citealt{gratton04} and references therein). 
However, it has been known for many years now that globular clusters exhibit 
star-to-star abundance variations for the light elements C, N, O, Na, Mg, 
and Al (e.g., see review by \citealt{kraft94}). Specifically, 
the abundances of C and O are low when N is high and anticorrelations are 
found between O and Na as well as Mg and Al. Recently, 
variations in the abundance of fluorine have been discovered in giants in 
M 4 where the amplitude of the dispersion exceeds that of O \citep{smith05}. 

It is generally assumed that the light element variations arise from 
proton-capture reactions (CNO-cycle, Ne-Na chain, and Mg-Al chain), though
the specific nucleosynthetic site(s) remain elusive. 
One possibility for the origin of the star-to-star abundance variations is
deep-mixing and internal nucleosynthesis within the observed stars. Evidence
for this ``evolutionary scenario'' include C and N abundances \citep{ss91} that 
vary with location on the red giant branch (RGB). Extensive mixing down to 
very hot layers is necessary to change the surface composition of Na, Mg, and
Al. Such mixing is not predicted by standard models and the proposed 
mechanisms include meridional circulation \citep{sm79}, turbulent 
diffusion \citep{charbonnel95}, and hydrogen-burning shell flashes  
\citep{fujimoto99,aikawa01,aikawa04}. An alternative possibility is 
pollution from intermediate-mass asymptotic giant branch stars (IM-AGBs), first 
suggested by \citet{cottrell81} to explain the Na and Al enhancements observed 
in CN strong stars in NGC 6752. In IM-AGBs, hydrogen-burning at the base of the
convective envelope, so-called hot bottom burning (HBB), can produce the 
observed C to Al abundance patterns. 
Either the ejecta from IM-AGBs pollute the proto-cluster 
gas from which the present cluster members form or the ejecta are accreted by 
present cluster members. The strongest evidence for this ``primordial 
scenario'' has come from observations of main-sequence stars in which abundance 
variations of O, Na, Mg, and Al have been found 
\citep{gratton01,ramirez03,cohen05}. In these unevolved stars, 
the internal temperatures are too low to run the Ne-Na or Mg-Al 
chains which therefore precludes internal mixing as a viable 
explanation for the star-to-star composition differences. 

In M 4, \citet{smith05} found that F varied from star-to-star and that the F
abundance was correlated with O and anticorrelated with Na and Al. Since
destruction of F is expected to take place 
during HBB in IM-AGBs \citep{lattanzio04}, the 
observed dispersion of F is in qualitative agreement with IM-AGBs 
being responsible for the globular cluster abundance anomalies. However, 
a quantitative test involving recent yields for AGB stars combined with a 
standard initial mass function showed that the observed abundance variations
cannot be reproduced via pollution from 
AGB stars \citep{fenner04}. \citet{denissenkov03} and \citet{denissenkov04}
also find flaws in the AGB pollution scenario based on calculated yields
from AGB models. \citet{ventura05} caution 
that theoretical yields from AGB models are critically dependent upon the 
assumed mass-loss rates and treatment of convection such that the 
predictive power of the current AGB models is diminished. 
That there is still no satisfactory explanation for the star-to-star abundance 
variations seen in every well studied Galactic globular cluster would suggest 
that our understanding of globular cluster chemical evolution and
stellar nucleosynthesis is incomplete. 

Two neutron-capture elements, rubidium and lead, may offer further clues  
regarding the processes that gave rise to the star-to-star abundance 
variations and possibly the formation of globular clusters. 
Rb has two stable isotopes, $^{85}$Rb and $^{87}$Rb. While the solar abundance of 
Rb is due to 50\% $s$-process and 50\% $r$-process \citep{burris00}, the 
abundance of Rb relative to nearby elements such as Sr, Y, and Zr offers an 
insight into the neutron density at the site of the $s$-process 
and therefore the mass of the AGB star due to the 10.7 yr half-life of
$^{85}$Kr (e.g., \citealt{tomkin83,tomkin99,lambert95,busso99,abia01}). 
Along the $s$-process path, Rb is 
preceded by Kr. The path enters at $^{80}$Kr and exits at either $^{85}$Kr 
or $^{87}$Kr with $^{85}$Kr providing the branching point. 
At low neutron densities ($N_n~\le~1 \times 10^8$ cm$^{-3}$), $^{85}$Kr
$\beta$-decays to the stable isotope $^{85}$Rb. At high neutron densities, 
$^{85}$Kr will capture neutrons to form $^{86}$Kr and then $^{87}$Kr which 
$\beta$-decays to $^{87}$Rb (effectively stable with a half-life of 
$4.7\times10^{10}$ yr). Clearly the isotopic mix of Rb depends upon the
neutron density. Unfortunately, stellar Rb isotope ratios 
cannot be measured \citep{lambert76}. 
In the presence of a steady flow along the $s$-process path, the density 
of a nuclide satisfies the condition $\sigma_iN_i \simeq$ constant, where 
$\sigma_i$ and $N_i$ are the cross-section and abundance of nuclide $i$ 
respectively. The neutron-capture cross-sections differ by a factor of 
10 between the two Rb isotopes ($\sigma_{87}~=~\sigma_{85}/10$). 
The $^{85}$Kr branch does not affect the Zr abundances since the low and high
neutron density $s$-process paths converge at Sr. Therefore, in a high
neutron density environment such as the helium intershell during a thermal 
pulse in IM-AGBs, the Rb abundance may increase by a factor of 10 relative
to nearby $s$-process elements such as Sr, Y, and Zr. In reality, the situation 
is slightly more complex since neutron capture on $^{84}$Kr leads to the ground
state of $^{85}$Kr as well as a short lived isomeric state that decays to 
either the $^{85}$Kr ground state or $^{85}$Rb (see \citealt{beer89} for more 
details). 
IM-AGBs of solar metallicity are expected to have a high neutron density with 
$^{22}$Ne($\alpha$,n)$^{25}$Mg providing the neutron source. (Low-mass AGBs, 
whose neutron source is $^{13}$C($\alpha$,n)$^{16}$O, provide 
a lower neutron density.) For metal-poor or 
zero metallicity IM-AGBs, \citet{busso01} suggest that such stars do run the 
$s$-process though the details are model dependent. 
Nevertheless, the Rb abundance relative to Sr, Y, and Zr (which are not 
affected by the $^{85}$Kr branch) is a potential diagnostic of the s-process 
site and may offer an additional insight into the role of IM-AGBs in the
chemical evolution of globular clusters. 

The isotopes of Pb, along with bismuth, comprise the last stable nuclei along 
the $s$-process path. In low-mass AGB stars, the neutron source is provided by
$^{13}$C 
whereas in IM-AGBs, $^{22}$Ne provides the neutron 
source with the division occurring at roughly 4 M$_\odot$. In low-mass AGBs 
and IM-AGBs of low metallicity, overabundances of Pb and Bi may be expected 
if the neutron supply per seed exceeds a certain 
value (e.g., \citealt{goriely00,travaglio01,busso01}). 
\citet{goriely01} suggest that for AGB stars with $Z < 0.001$, 
the available neutrons per seed nuclei is greater than the number required 
to produce Pb and Bi. 
\citet{travaglio01} suggest that metal-poor IM-AGBs play only a minor role in 
the production of Pb though for their 5 M$_\odot$ model, 
the Pb yields do not change between [Fe/H] = $-$1.3 and solar. 
\citet{herwig04} suggest that metal-poor IM-AGBs efficiently activate the
$^{22}$Ne neutron source though quantitative $s$-process yields are not
presented. \citet{busso01} predict 
high yields of Pb from metal-poor 
IM-AGBs. The ratio of Pb/La and Pb/Ba can be used to probe the nature of the
$s$-process in metal-poor AGB stars \citep{gallino98,goriely00}. 
Numerous observational studies have found considerable overabundances 
of Pb in stars that exhibit large $s$- and/or $r$-process enhancements 
(e.g., \citealt{cowan96,sneden00c,aoki00,aoki01,aoki02,vaneck01,vaneck03,johnson02a,lucatello03,sivarani04,ivans05} and references therein). If the globular cluster 
star-to-star abundance variations are due to pollution from metal-poor 
IM-AGBs, we may 
expect large overabundances of Pb and a dispersion in Pb abundances despite 
the absence of variations and excesses in other $s$-process elements. 

In this paper, we present measurements of Rb and Pb in the globular cluster
NGC 6752 as well as measurements of Pb in the globular cluster M 13. While
Rb has been measured in two globular clusters, 
$\omega$ Cen \citep{smith00} and NGC 3201 \citep{gonzalez98}, as far
as we are aware these are the first measurements of Pb in a globular cluster. 
We chose the globular clusters M 13 and NGC 6752 because they exhibit the 
largest amplitude for the Al variation of all the well studied Galactic
globular clusters and therefore offer the best opportunity to find abundance 
variations for Rb and Pb. Previous studies of M 13 include 
\citet{cohen78} and \citet{peterson80} who found large Na variations, 
\citet{shetrone96a,shetrone96b} who showed that the Mg isotope ratios were
not constant, 
\citet{cohen05} who discovered abundance variations in unevolved stars, as
well as analyses by \citet{kraft92,kraft97}, \citet{pila96}, and
\citet{sneden04a}. Previous studies of NGC 6752 include \citet{cottrell81} 
who discovered the Na and Al enhancements, \citet{ss91} who found C and N
to systematically vary according to evolutionary status, 
\citet{gratton01} and \citet{grundahl02} who discovered O-Al variations in 
unevolved stars, \citet{6752} (hereafter Y03) who found variations in Mg 
isotope ratios, \citet{67522} (hereafter Y05) who presented evidence for
slight abundance variations of Si, Y, Zr, and Ba, and \citet{pasquini05} 
who measured Li in main sequence stars.

\section{Observations and data reduction}
\label{sec:data}

The list of candidates included 5 giants in NGC 6752 previously studied in
Y03 and Y05, 4 giants in M 13 previously studied by 
\citet{shetrone96a,shetrone96b}, and the comparison star HD 141531, a giant
whose evolutionary status and stellar parameters are comparable to the cluster
giants. Though we were restricted to the brightest giants, the globular 
cluster stars were deliberately selected to span a large range 
of the star-to-star abundance variations. 
Table \ref{tab:param} contains the list of targets observed using 
either the Subaru or Magellan telescopes. 

Observations of the M 13 giants and the comparison field star HD 141531 were
obtained with the Subaru Telescope using the High 
Dispersion Spectrograph (HDS; \citealt{hds}) on 2004 June 1. A 0.4\arcsec\ 
slit was used providing a resolving power of 90,000 per 4 pixel resolution 
element with wavelength coverage from 4000~\AA~to 6700~\AA. 
For the Subaru data, one-dimensional wavelength calibrated normalized
spectra were produced in the standard way using the IRAF\footnote{IRAF 
(Image Reduction and Analysis
Facility) is distributed by the National Optical Astronomy
Observatory, which is operated by the Association of Universities
for Research in Astronomy, Inc., under contract with the National
Science Foundation.} package of programs.  

\begin{figure}
\epsscale{0.8}
\plotone{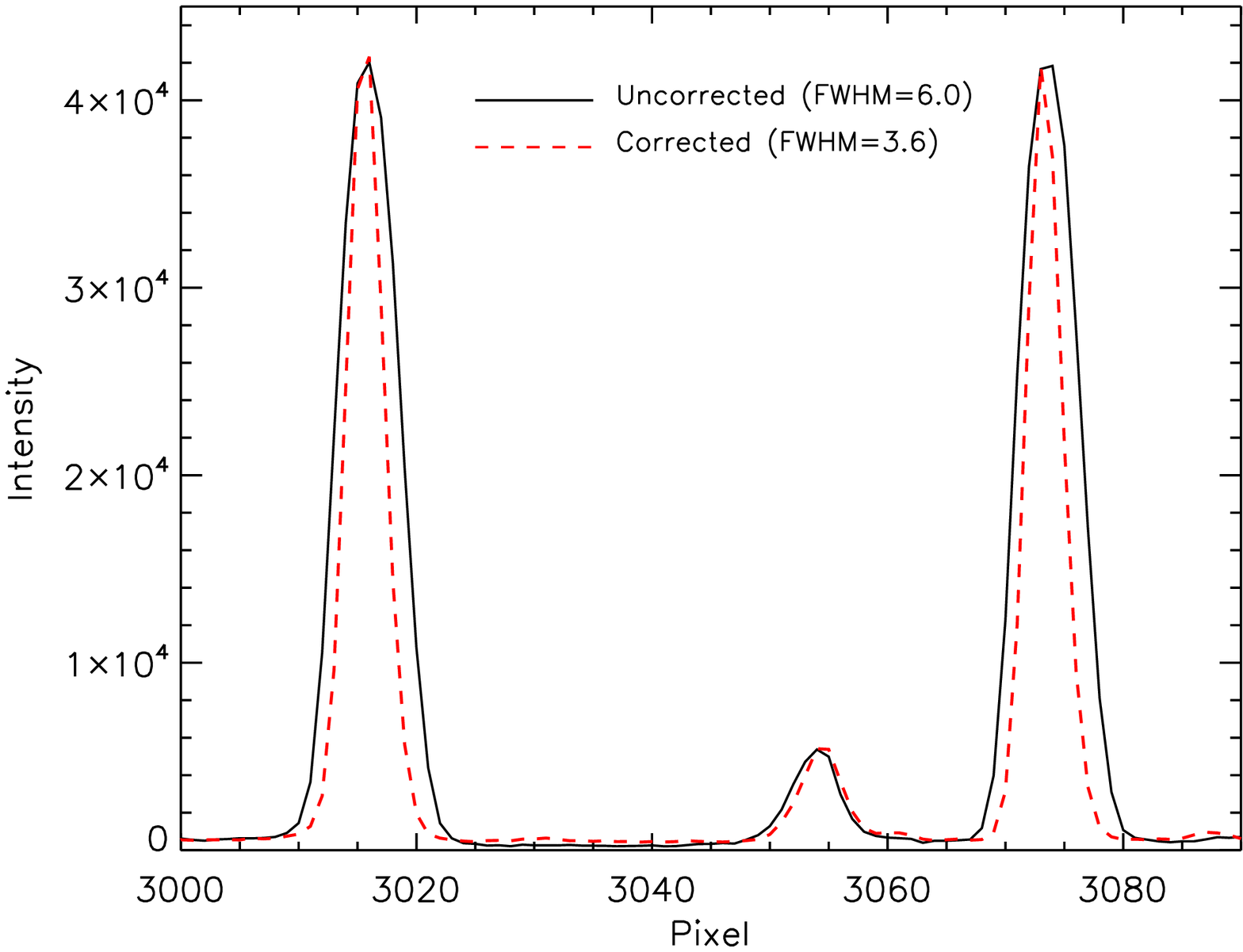}
\caption{Profiles of extracted ThAr lines from the same exposure taken
with Magellan-MIKE scaled to the same peak intensity. The solid
black line shows the ThAr lines extracted without taking into account the 
tilted slits. The dashed red line shows the ThAr lines extracted using the
{\sc mtools} package. 
Note how the FWHM
decreases when the correction is performed.\label{fig:tilt}}
\end{figure}

Observations of the NGC 6752 giants were obtained with the Magellan
Telescope using the Magellan Inamori Kyocera 
Echelle spectrograph (MIKE; \citealt{mike}) on 2004 April 3-5. A 0.35\arcsec\ 
slit was used providing a resolving power of 55,000 in the red and 65,000 in 
the blue per 4 pixel resolution element with wavelength coverage from
3800~\AA~to 8500~\AA. While IRAF was used for most of the data reduction,
extraction of the Magellan data must account for the 
``tilted'' slits, i.e., the lines are tilted with respect to the orders and 
the tilt varies across the CCD. While this is a feature of all cross-dispersed 
echelle spectrographs, for MIKE data the tilt is severe. 
We used the 
{\sc mtools}\footnote{http://www.lco.cl/lco/magellan/instruments/MIKE/reductions/mtools.html} 
set of tasks written by Jack Baldwin to correct for the tilt. Failure 
to make this correction would result in degradation of 
the spectral resolution as shown in Figure \ref{fig:tilt}. (The magnitude of this 
effect depends upon the aperture size applied to the order being extracted. 
For the stellar spectra, the decrease in spectral 
resolution would be smaller than for the 
Th-Ar comparison spectra by roughly a factor of 2.) 

\section{Analysis}
\label{sec:analysis}

\subsection{Stellar parameters and the iron abundance}
\label{sec:param}

\begin{figure}
\epsscale{0.8}
\plotone{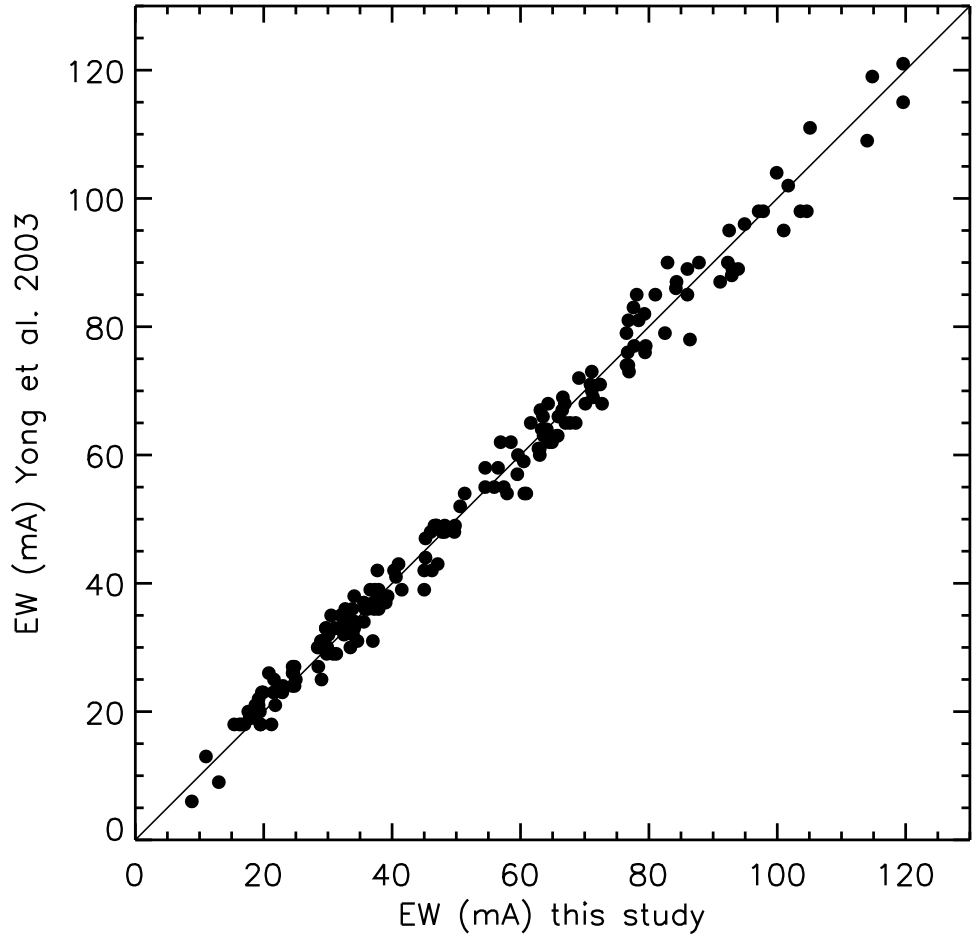}
\caption{Comparison of the equivalent widths of Fe I and Fe II lines 
between this study (Magellan data) and Y03 (VLT data) for the sample 
of NGC 6752 giants.\label{fig:ew_compm}}
\end{figure}

\begin{figure}
\epsscale{0.8}
\plotone{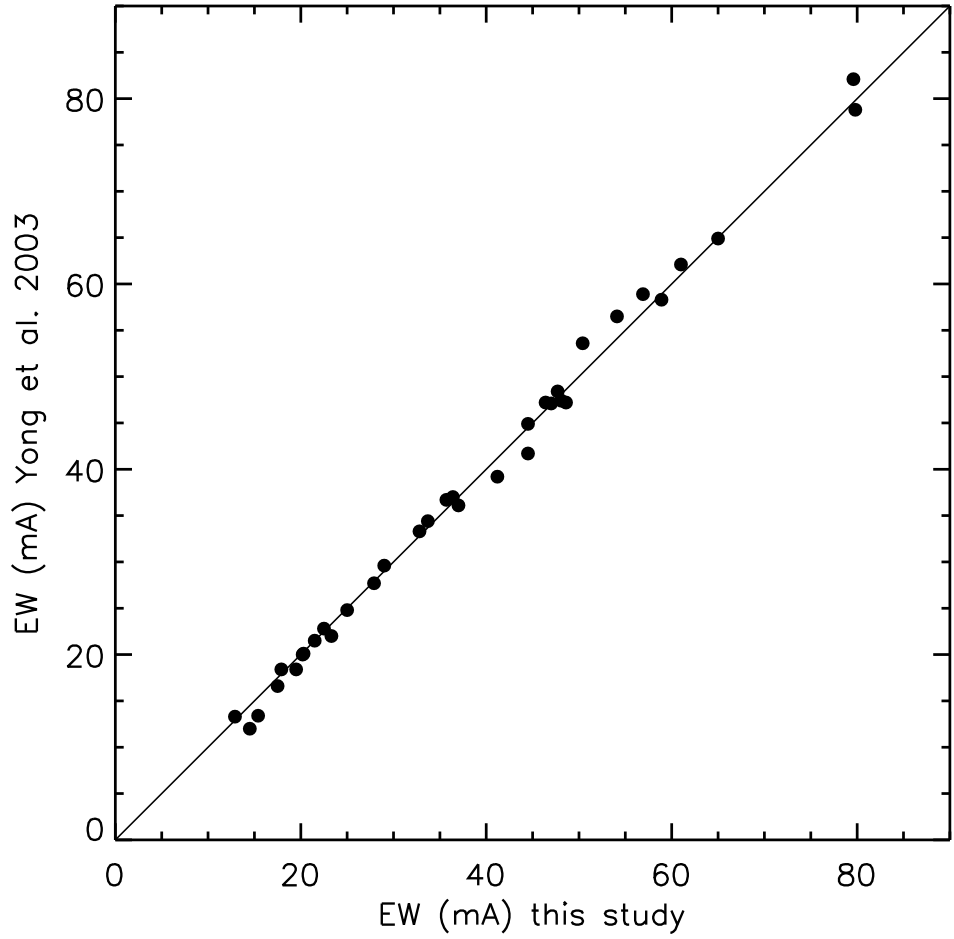}
\caption{Comparison of the equivalent widths of Fe I and Fe II lines 
between this study (Subaru data) and Y03 (VLT data) for
HD 141531.\label{fig:ew_comps}}
\end{figure}

The first step in the analysis was to determine the stellar parameters: 
the effective temperature (\teff), the surface gravity (log $g$), and
the microturbulent velocity ($\xi_t$). Equivalent widths (EWs) were measured
for a set of Fe\,{\sc i} and Fe\,{\sc ii} lines using routines in IRAF. 
We used the same set of Fe lines presented in Y03. In
Figure \ref{fig:ew_compm}, we compare the measured EWs for Fe\,{\sc i} and 
Fe\,{\sc ii} lines for the five NGC 6752 giants analyzed in Y03. The EWs 
measured in the Magellan data are in very good agreement with those measured 
in the VLT data. Therefore, for the five NGC 6752 giants, we adopt the same
stellar parameters used in Y03. 
In Figure \ref{fig:ew_comps}, we compare the EWs of Fe\,{\sc i} and Fe\,{\sc ii} 
lines for the comparison star HD 141531. The EWs measured in the Subaru data
are in very good agreement with those measured in the VLT data. For HD 141531,
we adopt the stellar parameters used in Y03. For the four M 13 giants, we 
determined the stellar parameters using spectroscopic criteria. As in Y03
and Y05, the model atmospheres were taken from the \citet{kurucz93} local 
thermodynamic equilibrium (LTE) stellar atmosphere grid and we used the 
LTE stellar line analysis program {\sc Moog} \citep{moog}. For the
effective temperature (\teff), we forced the Fe\,{\sc i} lines to show no
trend between abundance and lower excitation potential, i.e., excitation
equilibrium. To set the surface gravity ($\log g$), we forced the abundances
from Fe\,{\sc i} and Fe\,{\sc ii} to be equal, i.e., ionization equilibrium. 
We adjusted the microturbulent velocity ($\xi_t$) until there was no trend
between abundance and EW. The final [Fe/H] was taken to be the mean of all Fe
lines. Our stellar parameters for the M 13 giants compare very well with those
derived by \citet{shetrone96a,shetrone96b}, \citet{sneden04a}, and 
\citet{cohen05}. 

\subsection{Rubidium abundances}

\begin{figure}
\epsscale{0.8}
\plotone{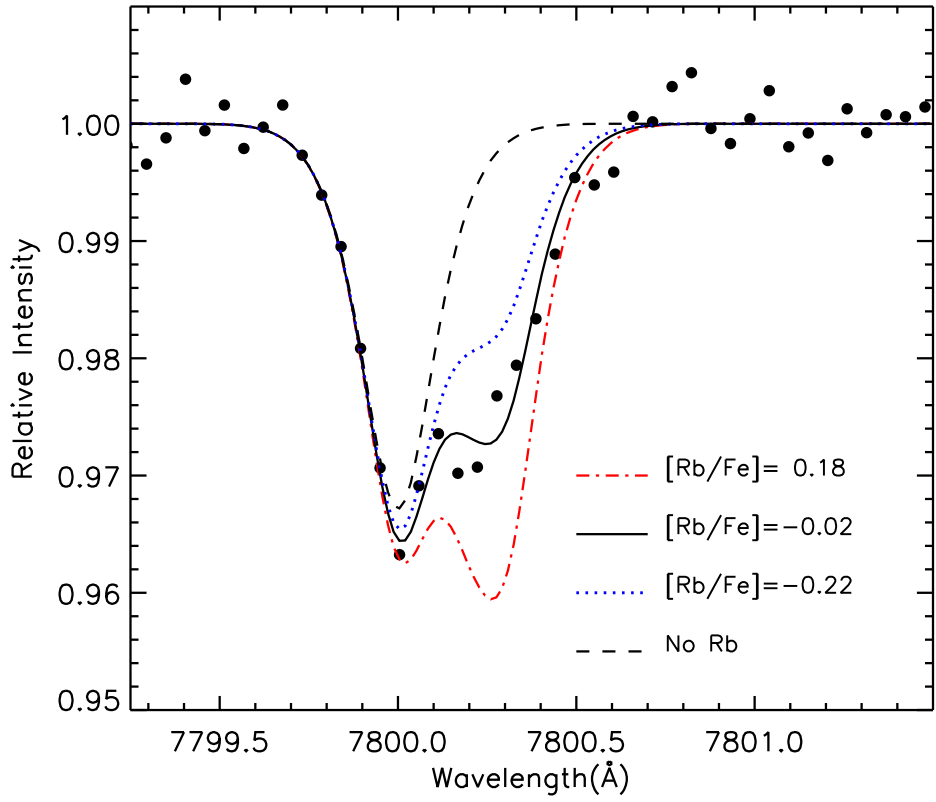}
\caption{Spectrum of NGC 6752 B702 near the 7800~\AA~Rb feature. Synthetic
spectra with differing Rb abundances are shown. 
\label{fig:b702_rb}}
\end{figure}

\begin{figure}
\epsscale{0.8}
\plotone{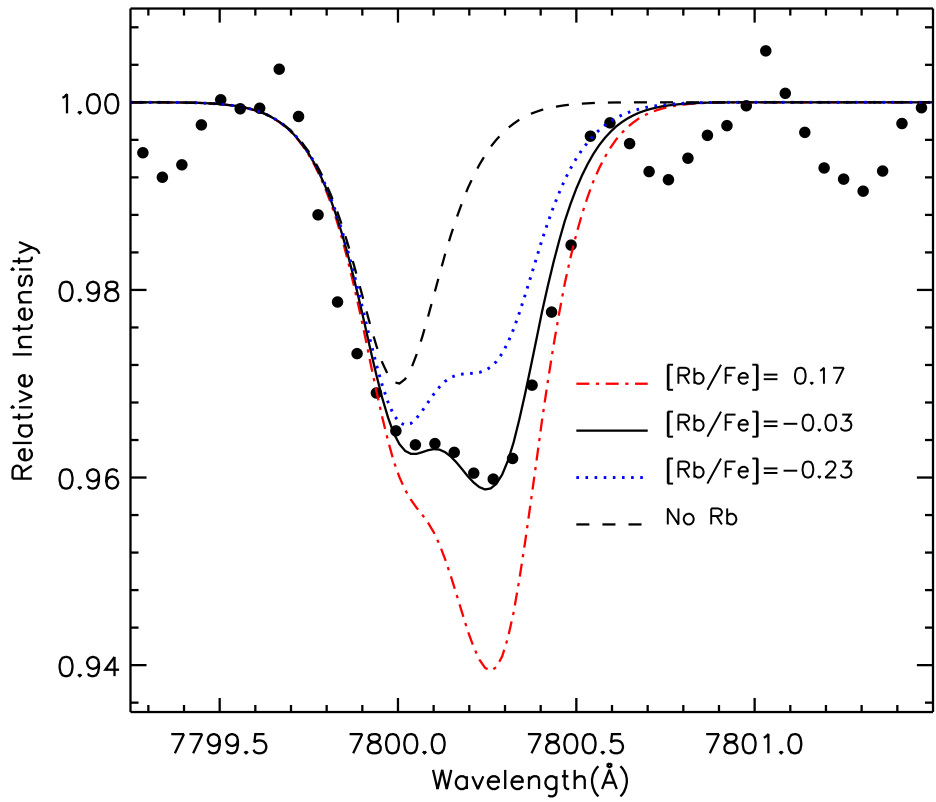}
\caption{Same as Figure \ref{fig:b702_rb} but for star NGC 6752 B1630.
\label{fig:mg1_rb}}
\end{figure}

For the M 13 giants, the HDS spectra did not incorporate the Rb line so we
were only able to measure Rb in the NGC 6752 giants. 
The abundances were determined via spectrum synthesis of the Rb\,{\sc i} 
line near 7800~\AA~(see Figures \ref{fig:b702_rb} and \ref{fig:mg1_rb}).  
Spectrum synthesis was essential for determining accurate abundances due to 
hyperfine splitting and isotopic shifts as well as blending from a stronger 
Si\,{\sc i} line. While the 7800~\AA~Rb\,{\sc i} line is only 3-4\% deep relative
to the continuum, the high quality spectra allow us to measure an abundance
from this line. Following \citet{tomkin99}, the wavelengths and relative
strengths for the isotopic and hyperfine structure components were taken from 
\citet{lambert76} and we assumed a solar isotope ratio of 
$^{85}$Rb/$^{87}$Rb = 3. The macroturbulent broadening was assumed to have
a Gaussian form and was estimated by
fitting the profile of the nearby Ni\,{\sc i} line at 7798~\AA. We then
generated synthetic spectra and varied the Si and Rb abundances to obtain
the best match to the observed spectrum. 
Ideally we would like to measure the Rb isotope ratio, $^{85}$Rb/$^{87}$Rb, 
but \citet{lambert76} were unable to measure accurate ratios in 
Arcturus even when using data with superior spectral resolution 
and signal-to-noise ratio (S/N) due to the presence of hyperfine structure 
and the small isotopic shift. While our 
tests confirmed that we could not measure accurate Rb isotope ratios, we
verify the finding by \citet{tomkin99} that the derived Rb abundances are 
not sensitive to the assumed isotope ratio. 
Using the \citet{kurucz84}
solar atlas, we measured an abundance log $\epsilon$(Rb) = 2.58 using a model
atmosphere with \teff/$\log g$/$\xi_t$ = 5770/4.44/0.85. Our derived solar 
Rb abundance is in very good agreement with the \citet{grevesse98} value, 
log $\epsilon$(Rb) = 2.60. 

\begin{figure}
\epsscale{0.8}
\plotone{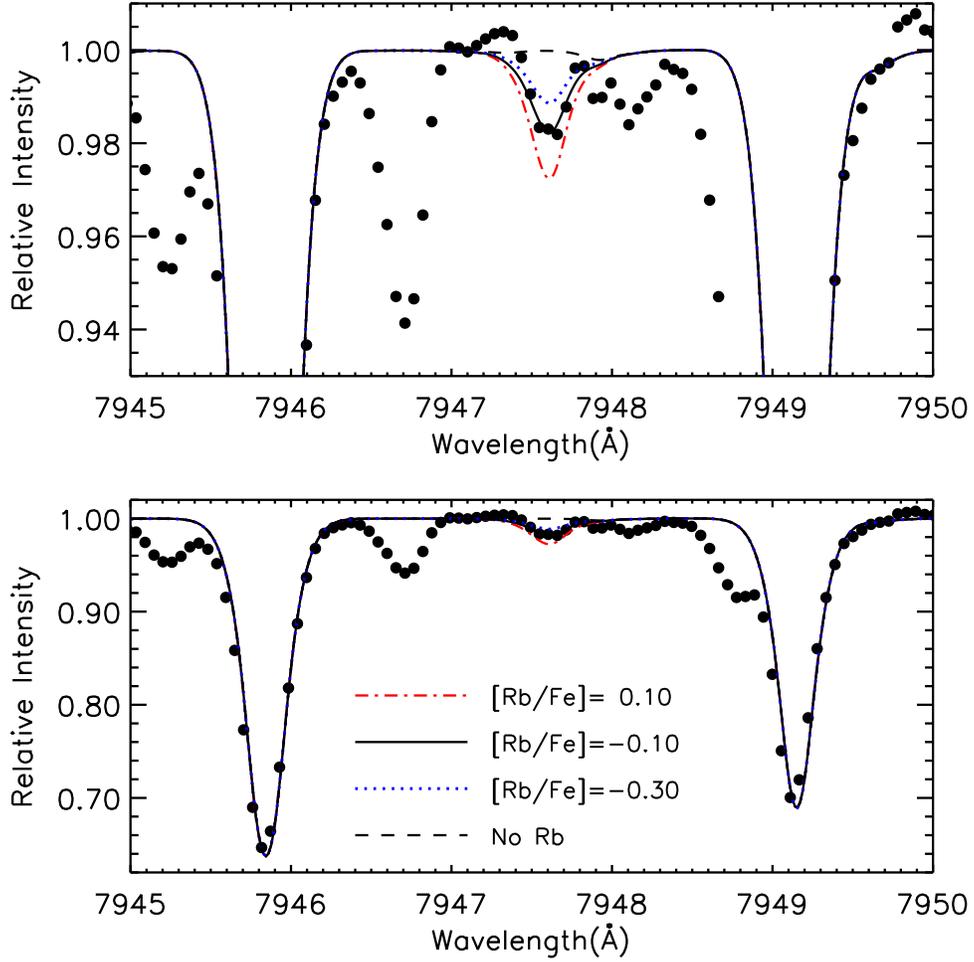}
\caption{ 
Spectrum of NGC 6752 B708 near the 7947~\AA~Rb feature. Synthetic
spectra with differing Rb abundances are shown. The lower panel is
identical to the upper panel except for the y-range. 
\label{fig:7947a}}
\end{figure}

The subordinate Rb\,{\sc i} line near 7947~\AA~is weaker by a factor of 2 and
is roughly 2\% deep relative to the continuum. We detect this line in all our 
spectra and preliminary analyses suggest that the abundances derived from this 
line agree with those from the 7800~\AA~line (see Figure \ref{fig:7947a}). 
However, we prefer to 
restrict our results to the 7800~\AA~line since the 7947~\AA~region is more 
crowded, i.e., the continuum placement strongly affects the derived Rb 
abundances from the 7947~\AA~line. Furthermore, 
unidentified blends, atmospheric absorption, 
and fringing are more prevalent in this spectral region. (CN lines lie in this
region may be absent in these metal-poor globular cluster stars.)

In the subsequent sections, we compare our globular cluster Rb abundances 
with field and cluster stars with [Fe/H] $> -2.0$ 
analyzed by other investigators. We 
now attempt to place the various Rb abundance measurements onto a common scale. 
While the isotope ratio of Rb cannot be measured in Arcturus, the elemental
abundance ratio is well known. Using the \citet{arcturus} Arcturus atlas,
we measured an abundance [Rb/H] = $-$0.55 using a model atmosphere with 
\teff/$\log g$/$\xi_t$ = 4300/1.5/1.55 obtained using the spectroscopic 
criteria described in Section \ref{sec:param}. The adopted Rb $gf$ value was 
identical to that used by \citet{tomkin99} and our derived Rb abundance is in 
very good agreement with their measured value, [Rb/H] = $-$0.58. 
Since Arcturus is common to both studies and our derived abundances 
are essentially identical, we therefore make no adjustment to 
the \citeauthor{tomkin99} Rb abundances. \citet{gratton94} use the same
$gf$ value, though the relative strengths of the hyperfine components
differ slightly. We do not adjust their Rb abundances. 
\citet{abia01} adopt a Rb $gf$ value
that differs from ours, so we adjust their Rb abundances 
by +0.08 dex. It is not clear what Rb $gf$ was used by \citet{gonzalez98}. 
Fortunately, Arcturus was also part of their sample. They derived an abundance
[Rb/H] = $-$0.45 and so we adjust their Rb abundances by $-$0.1 dex. 
\citet{smith00} find [Rb/H] = $-$0.52 and so we do not adjust their Rb 
abundances.  

\subsection{Lead abundances}

\begin{figure}
\epsscale{0.8}
\plotone{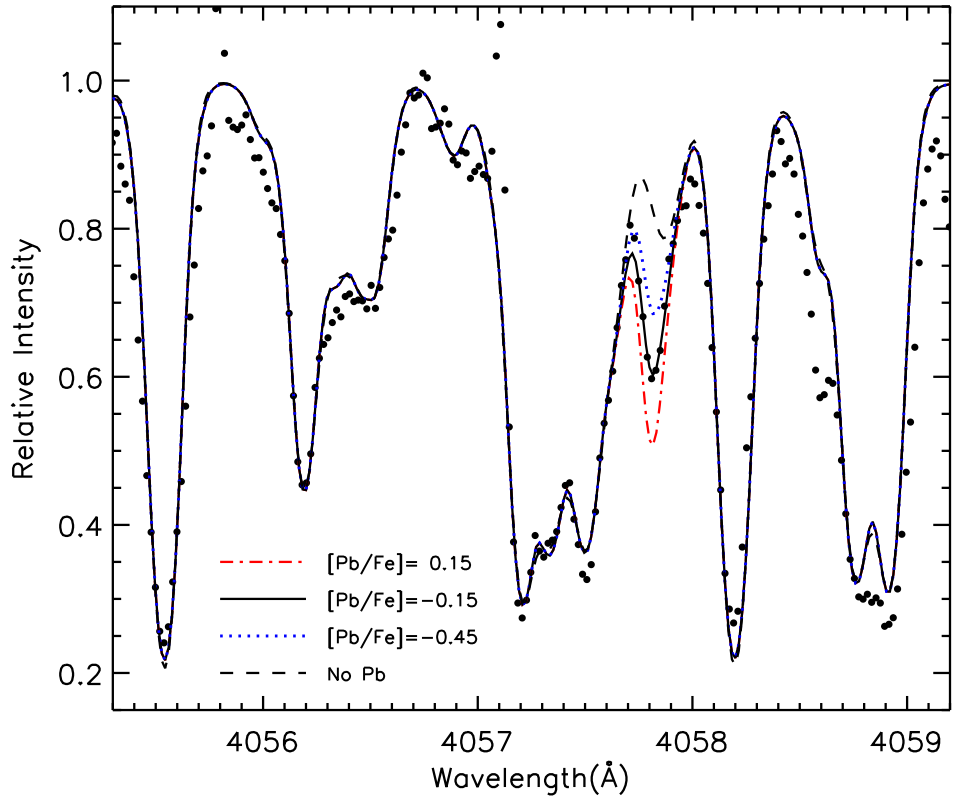}
\caption{Spectrum of NGC 6752 B708 near the 4058~\AA~Pb feature. Synthetic
spectra with differing Pb abundances are shown. 
\label{fig:b708_pb}}
\end{figure}

\begin{figure}
\epsscale{0.8}
\plotone{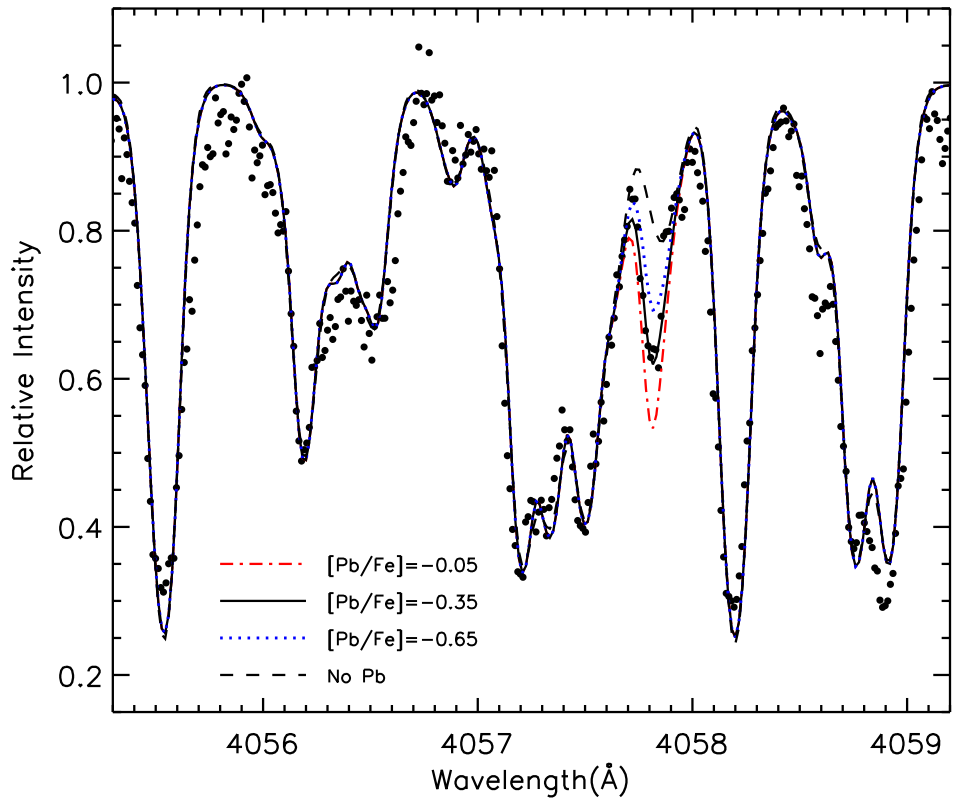}
\caption{Same as Figure \ref{fig:b708_pb} but for star M 13 L973.
\label{fig:m13_l973_pb}}
\end{figure}

The Pb abundances were determined via spectrum synthesis of the Pb\,{\sc i}
line near 4058~\AA~(see Figures \ref{fig:b708_pb} and \ref{fig:m13_l973_pb}). 
Abundances from the Pb line near 3683~\AA~could not be determined due to the
lack of flux in the blue 
for these cool giants. While the region centered near 4058~\AA~is crowded 
with molecular lines of CH as well as atomic lines from Mg, Ti, Mn, Fe, 
and Co, our syntheses provide a very good fit to the region 
demonstrating that reliable Pb abundances can be 
extracted. The macroturbulent broadening was estimated by fitting 
the profiles of the nearby lines. We adopted the same $gf$ value used
by \citet{aoki00,aoki01,aoki02}. Following \citet{aoki02}, our synthesis 
accounted for the hyperfine and isotopic splitting as well as the isotopic 
shifts. (The stable isotopes are $^{204}$Pb, $^{206}$Pb, $^{207}$Pb, and 
$^{208}$Pb.) We again assumed a solar isotope ratio for Pb though as with
Rb, our tests confirmed that the derived 
elemental Pb abundance was not sensitive to this choice. For the solar Pb
abundance, we adopted log $\epsilon$(Pb) = 1.95 from \citet{grevesse98}. 

In the subsequent sections, we compare the globular cluster Pb abundances
with those obtained in field stars with [Fe/H] $>-2.0$ by other
investigators. However, we further restrict the comparison by 
avoiding stars with known $s$-process enhancements leaving only a handful
of stars from two studies, \citet{sneden98} and \citet{travaglio01}. Again
we attempt to put the Pb abundance measurements onto a common scale. Our $gf$ 
value is identical to that used by \citet{sneden98} so we do not adjust
their Pb abundances. \citet{travaglio01} 
used a different Pb line (3683~\AA) and they did not list the adopted 
$gf$ value. Since there are no stars common to both analyses, we do not make 
an adjustment to their Pb abundances and caution that there may be a
systematic offset. 

\subsection{Additional elements}

We also measured abundances for Al, Si, Y, Zr, La, and Eu in the M 13 
giants and the 
comparison star HD 141531 using the same lines presented in Y05. These
measurements were performed to ensure that the abundances would be
on the same scale as the NGC 6752 giants studied in Y05. Zr 
was chosen because we compare the
Rb and Zr abundances to look for a large ratio [Rb/Zr] as well as a 
detectable dispersion, i.e., the hallmark of a high 
neutron-density environment and the possible signature of pollution from 
IM-AGBs. Al, Si, and Y were also chosen because Y05 found evidence 
for correlations between Al and Si, 
Al and Y, and Al and Zr in NGC 6752. While our sample size in M 13 is
small, it would be interesting to see if similar correlations are 
present. La and Eu were measured since these neutron-capture elements
offer an insight into the ratio of $s$-process to $r$-process material.
Furthermore, the ratio [Pb/La] has been used to test predictions from
AGB models. In Table \ref{tab:abund} we present our measured 
elemental abundances for Al, Si, Rb, Y, Zr, La, Eu, and Pb in the program stars. 
The adopted solar abundances for Al, Si, Y, Zr, La, and Eu were  
6.47, 7.55, 2.24, 2.60, 1.13, and 0.52 respectively. 

\begin{figure}
\epsscale{0.8}
\plotone{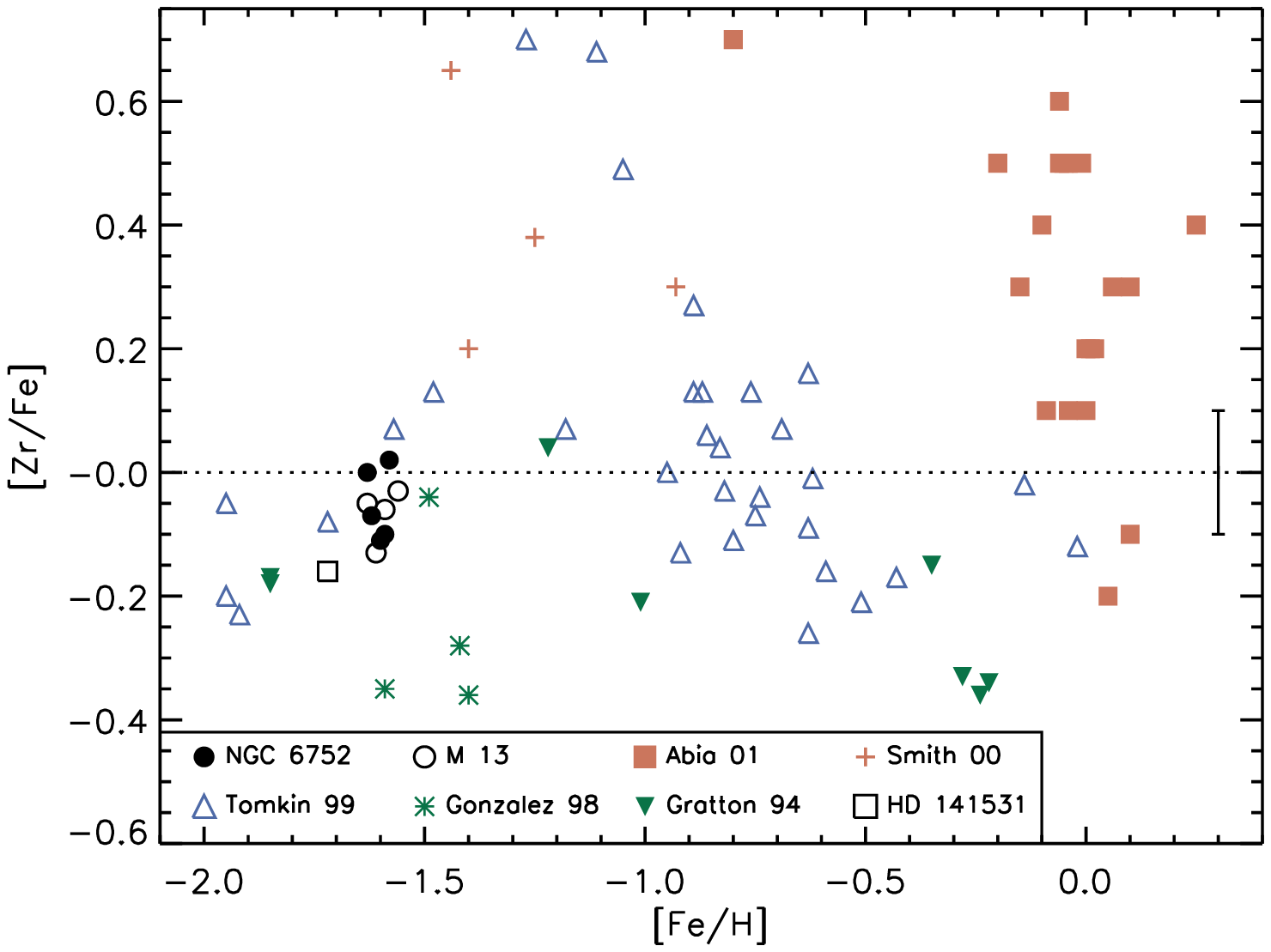}
\caption{[Zr/Fe] versus [Fe/H]. Closed black circles show our measurements for 
NGC 6752, the open black circles show M 13, the open black square shows
HD 141531, the closed green triangles are from \citet{gratton94}, the 
green asterisks represent data from \citet{gonzalez98}, 
open blue triangles are from \citet{tomkin99}, 
red plus signs represent data from \citet{smith00}, and 
filled red squares show data from \citet{abia01}. 
A representative error bar is shown and the Zr abundances have been
shifted onto the \citeauthor{smith00} scale (see text for details). 
\label{fig:zr}}
\end{figure}

We attempt once more to put the abundance measurements onto a common 
scale by considering the $gf$ values used by the various studies to which
we compare our abundances. The element we focus upon is Zr (in
order to compare [Rb/Zr] between the samples). We shift all the
Zr abundances onto the \citet{smith00} scale in order to compare
with their theoretical predictions for [Rb/Zr] 
from low and intermediate-mass AGBs (their 
Figure 14). (Our Rb abundances were already on the Smith scale.) 
Abundance measurements for Zr are complicated by the fact that
the laboratory Zr $gf$ values from \citet{zr} are smaller than the 
solar $gf$-values by 0.41 dex \citep{tomkin99}. We must therefore take care 
and account for both the adopted solar abundance and the $gf$ value. 
\citet{smith00} adopt the \citet{zr} $gf$-values and a solar abundance 
log $\epsilon_\odot$(Zr) = 2.90. 
We used the \citet{zr} $gf$-values and a solar abundance 
log $\epsilon_\odot$(Zr) = 2.60 \citep{grevesse98}. Therefore we adjust
our Zr abundances by $-$0.30 dex to ensure that we are on the same scale
as \citet{smith00}. 
Similarly, we adjust the Zr abundances of \citet{gratton94} and \citet{abia01} 
by $-$0.30 dex since they adopt the same $gf$ values and a very 
similar solar abundance used in our analysis. 
\citet{gonzalez98} adopt the \citet{grevesse98} solar abundance and a
different $gf$ value so we adjust their Zr abundances by +0.06 dex. 
\citet{tomkin99} adopt the \citet{grevesse98} solar abundance and a 
different $gf$ value so we adjust their Zr abundances by +0.11 dex. 
These adjustments are substantial.
When we compare the ratio [Rb/Zr], we also consider how the comparison 
would fare had we not made these abundance corrections. 
To assess the validity of these adjustments, we measured the 
Zr abundance for Arcturus and found [Zr/H] = $-$0.66. 
\citeauthor{smith00} measured [Zr/H] = $-$0.96, 
\citeauthor{gonzalez98} found $-$1.18, and \citeauthor{tomkin99} 
found $-$1.00. Therefore, applying the abundance 
corrections based on the $gf$ values and solar abundances 
ensures that the Zr abundances are on the \citet{smith00} scale, e.g., 
for Arcturus we find $-$0.96 (this study), 
$-$1.12 (\citeauthor{gonzalez98}), $-$0.89 (\citeauthor{tomkin99}), 
and $-$0.96 (\citeauthor{smith00}). 
Interestingly, if we had used the same solar abundance as \citet{smith00},
our [Zr/Fe] abundances in Y05 would have been closer to the [Y/Fe] values
and for M 13 we would have found [Y/Fe] $\simeq$ [Zr/Fe]. 
In Figure \ref{fig:zr}, we plot [Zr/Fe] versus [Fe/H] with the abundances
shifted to the \citet{smith00} scale. At the metallicity of NGC 6752 and M 13, 
the field and cluster stars have similar ratios of [Zr/Fe]. The comparison 
field star HD 141531 has [Zr/Fe] almost identical to the globular clusters.

As in Y03, we estimate the internal errors in the stellar parameters to be
\teff~$\pm $ 50K, $\log g~\pm$ 0.2, and $\xi_t~\pm$ 0.2. 
In Table \ref{tab:parvar}, we show the abundance dependences upon the model 
parameters. 

\section{Discussion}
\label{sec:discussion}

\subsection{Rubidium}

While our mean Rb abundance for NGC 6752 is [Rb/Fe] = $-$0.17 $\pm$ 0.06 
($\sigma$ = 0.14), the abundances appear to concentrate around two distinct 
values. There are two stars with [Rb/Fe] $\simeq$ $-$0.02 and three stars with
[Rb/Fe] $\simeq$ $-$0.25. The two stars with the higher Rb abundances do not
have the highest Al abundances and the three stars with the lower Rb abundances
are not exclusively the stars with the lowest Al abundances. Given the weakness
of the Rb line, the uncertainties in the derived Rb 
abundances (see Table \ref{tab:parvar}), and the small 
sample size, it is unlikely that the Rb abundances show a 
dispersion in NGC 6752. Nor do we find evidence for a correlation between 
[Al/Fe] and [Rb/Fe], though we recognize that our sample size (5 stars) 
is much more limited than in Y03 and Y05 (38 stars). 
Unfortunately, observations of M 13 and the 
comparison field star HD 141531 did not incorporate the Rb line. 

\begin{figure}
\epsscale{0.8}
\plotone{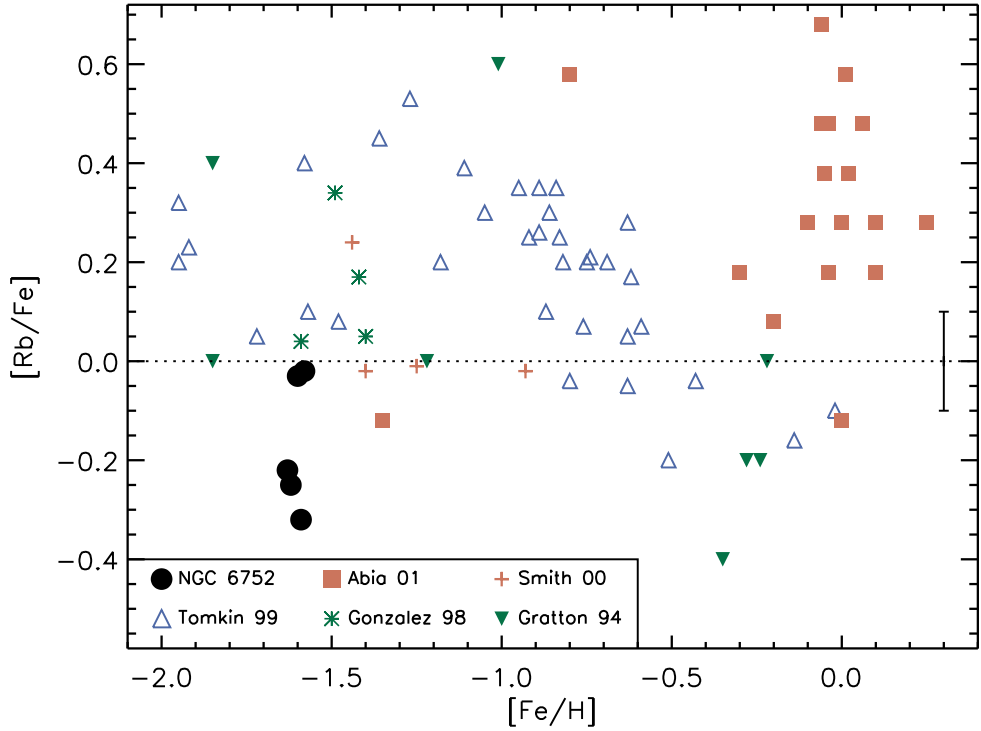}
\caption{[Rb/Fe] versus [Fe/H]. Closed black circles show our measurements for 
NGC 6752, the closed green triangles are from \citet{gratton94}, the 
green asterisks represent data from \citet{gonzalez98}, 
open blue triangles are from \citet{tomkin99}, 
red plus signs represent data from \citet{smith00}, and 
filled red squares show data from \citet{abia01}. 
A representative error bar is shown. 
\label{fig:rb}}
\end{figure}

In Figure \ref{fig:rb}, we compare our Rb abundances with those measured
in dwarfs and giants in the disk and halo (\citealt{gratton94} and 
\citealt{tomkin99}), globular cluster
giants (\citealt{gonzalez98} and \citealt{smith00}), 
and carbon-rich AGB stars \citep{abia01}. (While we retain their
stars in the plots, we note that the Abia 
sample contains very different objects that are difficult to 
analyze compared to the dwarfs and giants considered in the other studies.) 
Recall that we have made
small adjustments to the Rb abundances in an attempt to 
place them onto a common scale. At the metallicity of NGC 6752 
([Fe/H] = $-$1.6), our two stars in NGC 6752 with the highest [Rb/Fe] ratios 
have abundances compatible with the lower envelope of the \citet{tomkin99} 
sample. The two NGC 6752 stars also exhibit very similar abundances [Rb/Fe] 
to the \citet{gonzalez98} and \citet{smith00} globular cluster giants. 
Our three stars with the lower
[Rb/Fe] ratios appear unusual compared to the \citeauthor{tomkin99} 
sample. Only 1 star in the \citet{abia01} sample has [Fe/H] $< -1.0$ and it
is interesting that it has an abundance ratio [Rb/Fe] similar to those 
measured in NGC 6752. In general, Rb is not a well studied element and
the comparison data are limited. 

\begin{figure}
\epsscale{0.8}
\plotone{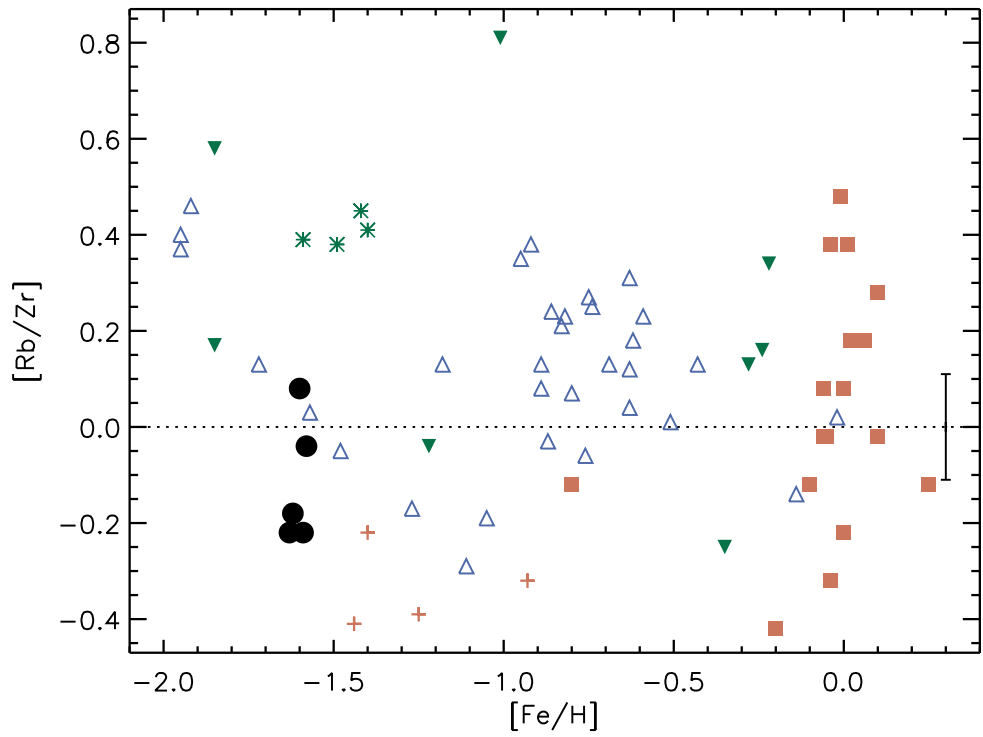}
\caption{[Rb/Zr] versus [Fe/H]. The symbols are the same 
as in Figure \ref{fig:rb}. Note that we shifted all 
abundances onto the \citet{smith00} scale (see text for
details).
\label{fig:rbzr}}
\end{figure}

For NGC 6752, we find a mean abundance [Rb/Zr] = $-$0.12 $\pm$ 0.06 
($\sigma$ = 0.13). (This abundance ratio has been shifted to the 
\citet{smith00} scale.) 
For the five NGC 6752 giants, the ratio [Rb/Zr] 
appears to show a dispersion. We suspect that this is attributable to
measurement uncertainties (primarily for Rb) rather than reflecting a  
real star-to-star scatter. We do not find a correlation between [Al/Fe] 
and [Rb/Zr]. 
In Figure \ref{fig:rbzr}, we compare the abundance ratio [Rb/Zr] between
NGC 6752 and various field and cluster stars. Note that in this Figure we have
shifted all abundances onto the \citet{smith00} scale since we will
utilize their theoretical predictions from low and intermediate-mass AGBs. 
At the metallicity of NGC 6752, we find that the two stars 
in NGC 6752 with the highest values of [Rb/Fe] also have the highest values of 
[Rb/Zr]. These two stars have similar [Rb/Zr] ratios to the \citet{gratton94} 
and \citet{tomkin99} samples at the same metallicity. While the $\omega$ Cen 
giants have abundance ratios [Rb/Zr] slightly lower 
than NGC 6752, this time the NGC 3201 giants appear to 
have much higher ratios of [Rb/Zr]. Note that the [Rb/Zr] ratios in NGC 3201
appear similar to the highest values seen in the \citet{tomkin99} sample. 
Unfortunately, the only star in the \citet{abia01} 
sample with [Fe/H] $< -1.0$ does not have a Zr measurement. However, it does
have [Rb/Sr] = $-$0.5 and [Rb/Y] = $-$0.6. If we assume for this star 
[Rb/Zr] = $<$[Rb/Sr],[Rb/Y]$>$ = $-$0.55, then the abundance is 
much lower than NGC 6752. 

For elements heavier than Si, globular clusters and field stars 
tend to have very similar 
abundance ratios [X/Fe] at a given [Fe/H] \citep{gratton04,sneden04}. 
Although the scatter is large and the sample sizes are limited, it would 
appear that cluster stars probably have similar, or perhaps slightly
lower abundance ratios of [Rb/Fe] and [Rb/Zr] compared to field
stars at a given [Fe/H]. 

Recall that we made substantial
adjustments to the Zr abundance. While consideration of the Arcturus Zr abundances 
would appear to validate this adjustment, we briefly consider how the 
comparison of [Rb/Zr] would have fared if these corrections were not applied. 
In this case, the ratio [Rb/Zr] would decrease by roughly 0.3 dex for 
NGC 6752 as well as for the \citet{gratton94} and \citet{abia01} samples. 
The \citet{tomkin99} and \citet{gonzalez98} samples would increase
by roughly 0.15 dex. 
NGC 6752 would therefore have unusually low ratios [Rb/Zr] compared to
field stars at the same metallicity. Similarly, the $\omega$ Cen compositions
would be unusually low though comparable to NGC 6752. 
The NGC 3201 giants would then have very high 
[Rb/Zr] ratios compared to other globular clusters and field stars at
the same metallicity. NGC 3201 is peculiar since it has a 
retrograde orbit and may have been a captured cluster \citep{vandenbergh93}. 
The capture hypothesis could not be demonstrated by \citet{gonzalez98} 
who found no unusual abundance ratios. 

\citet{smith00} compare [Rb/Zr] in $\omega$ Cen with predictions from AGB
models with various initial masses and initial metallicities 
(their Figure 14). Their Figure clearly shows how the ratio [Rb/Zr] can
vary by nearly a factor of 10 depending on whether a low-mass 
(1.5M$_\odot$) or high-mass (5M$_\odot$) AGB model is 
synthesizing the $s$-process elements. As anticipated from the
arguments given in Section \ref{sec:intro}, high-mass AGB models 
produce high [Rb/Zr] while low-mass AGB models produce low [Rb/Zr]. 
The magnitude of the difference in the predicted [Rb/Zr] between low-
and high-mass AGB models does not significantly 
change as the metallicity decreases from [Fe/H] = $-$0.5 to [Fe/H] = $-$2.0.
Comparing the observed abundances with the model predictions in 
\citeauthor{smith00} reveals that low-mass AGB stars (1-3 M$_\odot$) 
are responsible for the synthesis of the 
$s$-process elements in $\omega$ Cen. Inspection
of Figure 14 in \citet{smith00} shows that at the metallicity of NGC 6752,
[Fe/H] = $-$1.6, our measured ratio [Rb/Zr] = $-$0.12 is compatible with the 
$s$-process elements being synthesized in low-mass AGB stars though the assumed 
mass of the $^{13}$C pocket is critical. Predictions assuming a standard 
treatment for the $^{13}$C pocket or the $^{13}$C pocket increased by a factor
of 2 both suggest AGB stars with $<$ 3 M$_\odot$ are responsible for the
[Rb/Zr] ratios seen in NGC 6752. When the $^{13}$C pocket 
is diminished by a factor of 3, the AGB stars with masses $>$ 3 M$_\odot$
may explain the observed [Rb/Zr]. When
we return our Zr abundances to the original scale, [Rb/Zr] = $-$0.42,
the ratio in NGC 6752 
is only compatible with low-mass AGB stars. We note that the highest values
of [Rb/Zr] seen in the \citet{gratton94}, \citet{gonzalez98}, and 
\citet{tomkin99} samples all greatly exceed the 5 M$_\odot$ AGB model 
predictions. Such a discrepancy serves as a useful reminder of the unfortunate
reality that the detailed yields of $s$-process elements from AGB stars may be 
very model dependent \citep{busso01,ventura05}.

\subsection{Lead}

In NGC 6752, the mean Pb abundance is [Pb/Fe] = $-$0.17 $\pm$ 0.04 
($\sigma$ = 0.08) and in M 13 the mean abundance is 
[Pb/Fe] = $-$0.28 $\pm$ 0.03 ($\sigma$ = 0.06). 
Given the fairly large measurement uncertainty
for Pb (see Table \ref{tab:parvar}), neither 
NGC 6752 nor M 13 show any evidence for a dispersion in Pb abundances, though
our sample sizes for both clusters are small. Furthermore, the [Pb/Fe] 
ratios are very similar for these two clusters. As with Rb, there is no evidence
that the ratio [Pb/Fe] is correlated with [Al/Fe]. We note that one star, 
NGC 6752 PD1, has lower ratios of both [Rb/Fe] and [Pb/Fe] relative to other 
giants in this cluster. This subtle composition difference probably 
arises from uncertainties in the stellar parameters rather 
than representing a genuine difference. The comparison field giant 
HD 141531 has a ratio [Pb/Fe] essentially identical to the globular cluster
giants. 

\begin{figure}
\epsscale{0.8}
\plotone{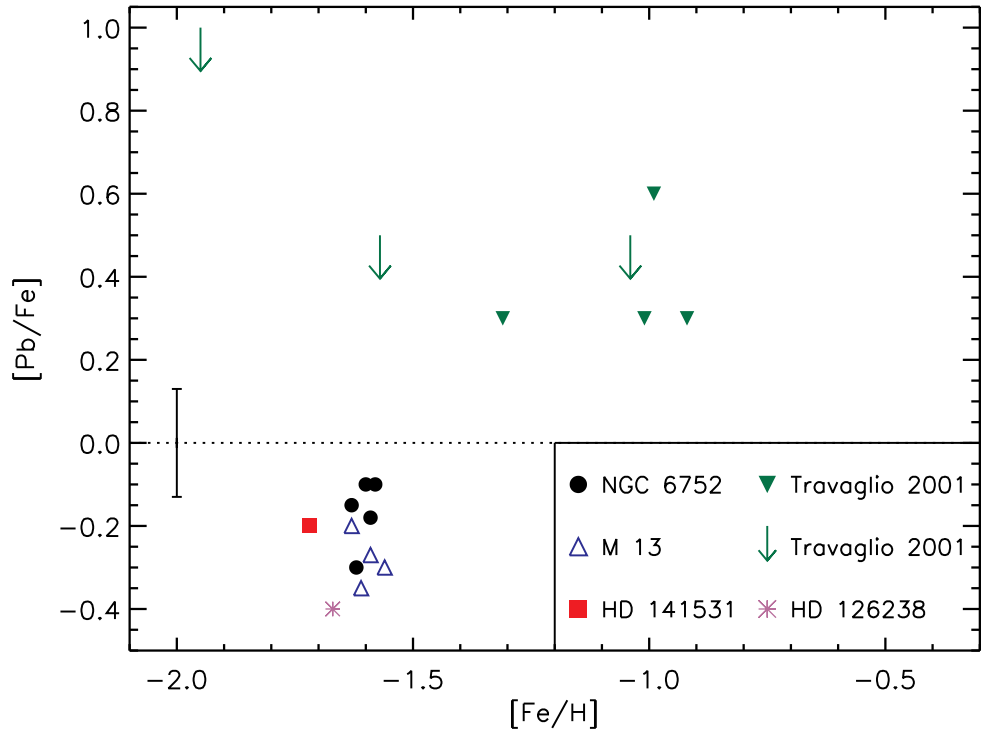}
\caption{[Pb/Fe] versus [Fe/H]. Closed black circles, open blue triangles, 
and the closed red square show our measurements for NGC 6752, M 13, and the
comparison field star HD 141531. The closed green triangles and green arrows
(upper limits) are from \citet{travaglio01} while the purple asterisk
shows HD 126238 from \citet{sneden98}. A representative error bar is shown. 
\label{fig:pb}}
\end{figure}

In Figure \ref{fig:pb}, we compare our Pb abundances with values
measured by \citet{sneden98} and \citet{travaglio01}. While Pb has been measured
in numerous stars with large $s$-process enhancements, it has been largely 
neglected in normal field stars presumably due to the difficulty of
the measurement. For HD 126238, the Pb abundance
measured by \citet{sneden98} is very similar to the globular cluster giants 
and HD 141531. The Pb abundances measured by \citet{travaglio01} in field
stars are larger than those measured in the globular clusters. The star
with [Pb/Fe] = 0.6 is a CH star with excess C and Ba and should not be 
considered a normal field star. 
Aside from the CH star, there are three stars with upper limits and another 
three Pb detections. Recall that there are no stars common to both studies 
and that the Pb $gf$ value was not published. \citet{travaglio01}
suggest that some of the Pb detections may be uncertain and therefore, the 
offset between the Pb abundances may be due to measurement errors and/or the 
adopted $gf$ value. 

\citet{travaglio01} not only measured Pb abundances in a handful of stars,
but they calculated the Galactic chemical evolution of Pb from a detailed
model. In their Figure 4, they plot the expected run of [Pb/Fe] versus 
[Fe/H] for the halo, thick disk, and thin disk. 
Since their prediction integrates over all AGB masses, 
it would be useful to learn how the predicted curve would differ (if at all) 
if the calculation was performed using low-mass or high-mass AGB models
exclusively as \citet{smith00} have done. At [Fe/H] = $-$1.6, 
the \citet{travaglio01} model predicts an abundance ratio 
[Pb/Fe] $\simeq$ $-$0.1. This prediction is in very good agreement with the
values measured in M 13, NGC 6752, HD 126238, and HD 141531. This agreement may
be regarded as evidence that globular cluster stars have virtually identical
Pb abundances as normal field stars. 

In normal field stars, Pb has been less 
studied than Rb. Clearly, it would be of great
interest to have additional Pb measurements in field and cluster stars. 
In Figures \ref{fig:b708_pb} and \ref{fig:m13_l973_pb}, our 
syntheses indicate that for cool giants in the metallicity regime 
$-$2.0 $<$ [Fe/H] $<$ $-$1.0, reliable Pb abundances can be measured 
even in stars that do not have large Pb or $s$-process enhancements. 

The ratio [Pb/La] may offer further clues regarding the nature of the
$s$-process in the AGB stars. \citet{vaneck03} found some stars with ratios 
of [Pb/La] $>$ +1.5, in agreement with predictions from metal-poor AGB
models \citep{gallino98,goriely00}. However, \citet{aoki02} and 
\citet{vaneck03} also found a large spread in the ratio [Pb/Ba]. In some
stars, the ratio [Pb/Ba] was sub-solar. Our mean ratio [Pb/La] for 
M 13 is $-$0.36 $\pm$ 0.05 ($\sigma$ = 0.10). For NGC 6572, our mean ratio
[Pb/La] = $-$0.23 $\pm$ 0.04 ($\sigma$ = 0.09). In both clusters, the 
mean ratios are similar and we note that they are both sub-solar and 
comparable to the lowest ratios seen in the \citet{vaneck03} sample. 
Curiously the subsample in \citet{vaneck03} with [Pb/La] $<$ 0 had 
extreme enhancements for [Pb/Fe] and [La/Fe]. 
The comparison field star HD 141531, has [Pb/La] = $-$0.20 which is
similar to the value seen in the globular clusters. 

\subsection{Additional elements}

While our sample in M 13 consists of 
only 4 stars, they span the extremities
of the Al variation. As in Y05, we again find that the most Al-rich stars may 
also exhibit slightly higher Si abundances than the most Al-poor stars. Further 
measurements of Al and Si in a large sample of stars in M 13 would be of
great interest to verify whether the correlation between Al and Si seen in 
NGC 6752 (Y05) is also present in M 13. The correlations between
Al and Y as well as Al and Zr found in NGC 6752 do not appear to be present
in the small M 13 sample. \citet{cohen05} measured abundances in 25 stars in
M 13 from the main sequence turn-off to the tip of the RGB. They were
unable to measure Al in most stars. When we consider their derived abundances, 
there appears to be an anticorrelation 
between O and Si as well as O and Y. Though the anticorrelation is driven
primarily by the one star with unusually low [O/Fe], such trends are intriguing 
and warrant further investigation. 

As noted in previous investigations of these clusters, the 
ratio of $s$-process to $r$-process material, [La/Eu], is sub-solar 
but greater than the scaled solar pure $r$-process value. For NGC 6752 
and M 13, the observed ratios of [La/Eu] show that 
AGBs have contributed to their chemical evolution. 
The ratio of La/Eu
in HD 141531 again confirms that it is a normal field star. 

\subsection{Consequences for the IM-AGB pollution scenario}

In Y03, we measured Mg isotope ratios in bright giants in NGC 6752. We found 
that the ratio varied from star-to-star. Specifically, $^{24}$Mg was 
anticorrelated with Al, $^{26}$Mg was correlated with Al, and $^{25}$Mg was 
not correlated with Al. As previously seen by \citet{shetrone96b} in M 13, 
these isotope ratios reveal that the Al enhancements result from proton capture 
on the abundant $^{24}$Mg. Proton capture on $^{24}$Mg within the Mg-Al chain is
predicted to only occur in AGB stars of the highest 
mass at their maximum luminosity \citep{karakas03}. 
So we suggested that the abundance variations
were due to differing degrees of pollution by IM-AGBs, an idea originally 
proposed by \citet{cottrell81}. These IM-AGBs must have the same iron abundance
as the present generation of cluster stars otherwise there would also be
a star-to-star abundance variation of Fe. (The same argument applies to whatever
stars are believed to be the source of the pollutants.) 

The Mg isotope ratios presented in Y03 offered further clues 
to globular cluster chemical evolution. At one extreme of the
abundance variation are cluster stars with O, Na, Mg, and Al compositions 
in accord with field stars at the same metallicity. We called such stars
``normal'' in anticipation that proton capture nucleosynthesis can 
produce O-poor, Na-rich, Mg-poor, and Al-rich material. 
At the other extreme of the abundance variations are the stars 
with high Na, high Al, low O, and low Mg. We referred to these stars as 
``polluted''. The pollution may have occurred via either the evolutionary or 
primordial scenario. In ``normal'' stars, we found ratios $^{25}$Mg/$^{24}$Mg
and $^{26}$Mg/$^{24}$Mg that exceeded field stars at the same metallicity
\citep{mghdwarf}. Of equal importance was the 
fact that these isotope ratios greatly exceeded 
predictions from metal-poor supernovae. We therefore suggested that these
unusually high isotope ratios could be explained if a 
previous generation of IM-AGBs of the highest mass polluted 
the natal cloud from which the cluster formed. The ejecta from this previous
generation must have been thoroughly mixed before the present generation
of stars began to form. This previous generation of IM-AGBs are probably
responsible for much of the Na, Al, and N as well as $^{25}$Mg and $^{26}$Mg. 

Our working hypothesis is that IM-AGBs played two crucial roles in globular 
cluster chemical evolution. 
Firstly, a prior generation of very metal-poor IM-AGBs are required
to produce the high $^{25}$Mg/$^{24}$Mg and $^{26}$Mg/$^{24}$Mg seen in
``normal'' stars. Secondly, a generation of 
IM-AGBs with the same Fe abundance as the
present cluster members pollutes the cluster environment. Differing degrees 
of pollution of natal clouds then produce the star-to-star abundance 
variations. The dispersion in the F abundances and the correlation between
F and O in M 4 \citep{smith05} appear to confirm the role of 
IM-AGBs in producing the abundance variations. 
However, not all the abundance patterns observed in globular clusters 
can be matched by the current theoretical yields from IM-AGBs 
\citep{denissenkov03,denissenkov04} nor can the abundance patterns 
be reproduced by chemical evolution models \citep{fenner04}. 

From a qualitative viewpoint, metal-poor IM-AGBs may produce $s$-process 
elements via the $^{22}$Ne neutron source. If activated, the $^{22}$Ne neutron source 
produces large amounts of Rb/Zr due to a critical branching point at 
$^{85}$Kr as described earlier. Theoretical models by \citet{busso01} suggest
that metal-poor IM-AGBs do run the $s$-process though the specific yields
depend on the details.   

Similarly, a qualitative assessment suggests that metal-poor IM-AGBs will 
produce Bi and Pb if the neutrons per seed nuclei exceed a certain value. 
In this case, Bi and Pb may show large enhancements with other $s$-process
elements showing only modest overabundances. Again, theoretical models 
can be found in which metal-poor IM-AGBs do produce lead (e.g., 
\citealt{goriely01} and \citealt{busso01}). 

In M 13 and NGC 6752, we did not find high ratios of [Rb/Fe], 
[Rb/Zr], or [Pb/Fe] compared with field stars at the same metallicity. 
If metal-poor IM-AGBs are responsible for the globular cluster star-to-star 
abundance variations, then our measurements strongly suggest that such stars
do not synthesize significant quantities of Rb or Pb. Also, if metal-poor 
IM-AGBs are responsible for the large abundances of $^{25}$Mg and $^{26}$Mg 
in ``normal'' cluster stars, then they do not synthesize Rb or Pb. 
Alternatively, if metal-poor IM-AGBs do synthesize significant quantities of
Rb and Pb, then they cannot be responsible for the abundance anomalies 
seen in globular clusters. 

Of course the possibility remains that the predicted yields of Rb and Pb from 
IM-AGBs are unreliable and/or model dependent (e.g., \citealt{ventura05}). 
It has been suggested that metal-poor IM-AGBs 
will not produce any $s$-process elements. As the mass of the AGB star 
increases, the size of the He intershell region decreases as does the duration of 
the thermal pulse \citep{lattanzio04}. Detailed predictions of the yields from 
metal-poor IM-AGBs by independent groups are required. Our observed Rb and Pb 
abundances may serve to constrain these models. 

Finally, we note that Rb and Zr are synthesized in IM-AGBs as well 
as via the weak $s$-process, 
i.e., He core burning in massive stars. 
However, \citet{travaglio04} suggest that the weak
$s$-process does not contribute to Zr
but a lighter element primary process in massive stars may be responsible 
for up to 18\% of the solar abundance of Zr.
\citet{chieffi04} present detailed 
yields from massive stars for a range of metallicities and masses. At 
$Z$ = 0, massive stars are predicted to produce low ratios, [Rb/Zr] = 
$-$0.5 to $-$2.4. If IM-AGBs have contributed to the chemical evolution
of NGC 6752, they would increase the ratios of [Rb/Zr] above those
produced by the supernovae. If IM-AGBs have
not played a role in the chemical evolution, perhaps we can use [Rb/Zr]
to probe the mass and metallicity range of the previous generation
of massive stars. 
For $Z$ = 10$^{-4}$, the predicted yields are independent of mass with 
[Rb/Zr] = $-$0.23. However, we note that for other metallicities, 
there is a mass-metallicity degeneracy
for the [Rb/Zr] yields that limits their use in probing the previous
generation of supernovae. 

\section{Concluding remarks}
\label{sec:summary}

We show for the first time the uniformity of the neutron-capture elements 
Rb and Pb in NGC 6752 and M13, the two globular clusters 
that exhibit the largest dispersion for Al. We also find the ratio [Rb/Zr]
to be constant. None of the abundance ratios [Rb/Fe], [Rb/Zr], or [Pb/Fe] 
are correlated with [Al/Fe] and the Rb and Pb abundances show sub-solar 
ratios [X/Fe]. If metal-poor IM-AGBs produce large amounts of
Pb and Rb as well as high ratios of [Rb/Zr], then such stars are not
responsible for the abundance variations, a conclusion already suggested 
by \citet{denissenkov03}, \citet{denissenkov04}, and \citet{fenner04}. 
If metal-poor IM-AGBs are responsible for the abundance variations, then
they cannot produce overabundances of Rb or Pb. 

For elements heavier than Al,
previous studies have shown that field and cluster stars generally have the 
same abundance ratios [X/Fe] at a given [Fe/H]. While our sample size is small
and the data for comparison field stars are limited, the two clusters 
we have studied have Rb abundance ratios [Rb/Fe] and [Rb/Zr] in 
reasonable agreement with the general field population (the clusters
may have slightly lower ratios). The Pb abundance ratio
[Pb/Fe] in globular clusters is in very good agreement with the limited sample
of field stars. At the metallicity of M 13 and NGC 6752, their Pb abundances 
are well matched by the predictions from the 
chemical evolution model by \citet{travaglio01}. 
In order to further our understanding of 
stellar nucleosynthesis and the chemical evolution of field and cluster stars, 
additional measurements of Rb and Pb in normal stars and globular 
clusters are welcomed as are further theoretical efforts to calculate
the yields from metal-poor IM-AGBs.  

\acknowledgments

This research has made use of the SIMBAD database,
operated at CDS, Strasbourg, France and
NASA's Astrophysics Data System. DY thanks John Lattanzio and Roberto Gallino 
for helpful discussions, Chris Sneden for providing a linelist for the Pb
region, and Bruce Carney for a thorough review of a draft of this paper. 
This research was performed while DBP held a National Research Council
Research Associateship Award at NASA's Goddard Space Flight Center.
DLL acknowledges support from the Robert A.\ Welch Foundation of Houston, Texas. 
This research was 
supported in part by NASA through the American Astronomical Society's Small 
Research Grant Program.

\clearpage

\clearpage

\clearpage

\clearpage

\clearpage

\clearpage

\clearpage

\clearpage

\clearpage

\clearpage

\clearpage

\clearpage

\begin{deluxetable}{lcccccccc} 
\tabletypesize{\footnotesize}
\tablecolumns{9} 
\tablewidth{0pc} 
\tablecaption{Exposure times and stellar parameters.\label{tab:param}}
\tablehead{ 
\colhead{Star} &
\colhead{Telescope} &
\colhead{Exposure} &
\colhead{S/N\tablenotemark{a}} &
\colhead{S/N\tablenotemark{a}} &
\colhead{\teff} &
\colhead{log $g$} &
\colhead{$\xi_t$} &
\colhead{[Fe/H]} \\
\colhead{} &
\colhead{} &
\colhead{Time (min)} &
\colhead{4050\AA} &
\colhead{7800\AA} &
\colhead{K} &
\colhead{} &
\colhead{km s$^{-1}$} &
\colhead{}
}
\startdata
M 13 L598 & Subaru & 60 & 39 & \ldots & 3900 & 0.00 & 2.25 & $-$1.56 \\
M 13 L629 & Subaru & 50 & 37 & \ldots & 3950 & 0.20 & 2.25 & $-$1.63 \\
M 13 L70\tablenotemark{b} & Subaru & 60 & 39 & \ldots & 3950 & 0.30 & 2.25 & $-$1.59 \\
M 13 L973\tablenotemark{c} & Subaru & 64 & 36 & \ldots & 3920 & 0.30 & 2.35 & $-$1.61 \\
NGC 6752 B702 & Magellan & 20 & 42 & 197 & 4050 & 0.50 & 2.10 & $-$1.58 \\
NGC 6752 B708 & Magellan & 50 & 55 & 402 & 4050 & 0.25 & 2.20 & $-$1.63 \\
NGC 6752 PD1 & Magellan & 26 & 47 & 197 & 3928 & 0.26 & 2.70 & $-$1.62 \\
NGC 6752 B1630 & Magellan & 27 & 44 & 247 & 3900 & 0.24 & 2.70 & $-$1.60 \\
NGC 6752 B3589 & Magellan & 21 & 46 & 240 & 3894 & 0.33 & 2.50 & $-$1.59 \\
HD 141531 & Subaru & 10 & 85 & \ldots & 4273 & 0.80 & 1.90 & $-$1.72 \\
\enddata

\tablenotetext{a}{S/N values are per pixel.}
\tablenotetext{b}{Alternative name II-67.}
\tablenotetext{c}{Alternative name I-48.}

\end{deluxetable}

\begin{deluxetable}{lcccrcrccc} 
\tabletypesize{\scriptsize}
\tablewidth{0pc} 
\tablecaption{Elemental abundances.\label{tab:abund}}
\tablehead{ 
\colhead{Star} &
\colhead{[Al/Fe]} &
\colhead{[Si/Fe]} &
\colhead{[Rb/Fe]} &
\colhead{[Y/Fe]} &
\colhead{[Zr/Fe]\tablenotemark{a}} &
\colhead{[Zr/Fe]\tablenotemark{b}} &
\colhead{[La/Fe]} &
\colhead{[Eu/Fe]} &
\colhead{[Pb/Fe]}
}
\startdata
M 13 L598 & 0.24 & 0.23 & \ldots & $-$0.10 & 0.27 & $-$0.03 & 0.00 & 0.40 & $-$0.30 \\
M 13 L629 & 0.74 & 0.31 & \ldots & $-$0.05 & 0.25 & $-$0.05 & 0.06 & 0.44 & $-$0.20 \\
M 13 L70 & 1.27 & 0.33 & \ldots & $-$0.12 & 0.24 & $-$0.06 & 0.14 & 0.43 & $-$0.27 \\
M 13 L973 & 1.28 & 0.35 & \ldots & $-$0.05 & 0.17 & $-$0.13 & 0.12 & 0.52 & $-$0.35 \\
NGC 6752 B702 & 1.04 & 0.43 & $-$0.02 & 0.04 & 0.32 & 0.02 & 0.09 & 0.28 & $-$0.10 \\
NGC 6752 B708 & 1.23 & 0.35 & $-$0.22 & 0.04 & 0.30 & 0.00 & 0.03 & 0.28 & $-$0.15 \\
NGC 6752 PD1 & 1.08 & 0.38 & $-$0.25 & 0.07 & 0.23 & $-$0.07 & 0.07 & 0.32 & $-$0.30 \\
NGC 6752 B1630 & 0.82 & 0.41 & $-$0.03 & 0.04 & 0.19 & $-$0.11 & 0.05 & 0.34 & $-$0.10 \\
NGC 6752 B3589 & 0.77 & 0.43 & $-$0.32 & 0.10 & 0.20 & $-$0.10 & 0.08 & 0.35 & $-$0.18 \\
HD 141531 & 0.01 & 0.23 & \ldots & $-$0.13 & 0.14 & $-$0.16 & 0.00 & 0.33 & $-$0.20 \\
\enddata

\tablenotetext{a}{Zr abundances using the $gf$ values and solar abundance
assumed in Y05.}
\tablenotetext{b}{Zr abundances when shifted onto the \citet{smith00} scale.}

\end{deluxetable}

\begin{deluxetable}{lrrr} 
\tabletypesize{\footnotesize}
\tablecolumns{4} 
\tablewidth{0pc} 
\tablecaption{Abundance dependences on model parameters for
NGC 6752 B1630.\label{tab:parvar}}
\tablehead{ 
\colhead{Species} &
\colhead{\teff~+ 50K} &
\colhead{$\log g$ + 0.2cgs} &
\colhead{$\xi_t$ + 0.2km s$^{-1}$}
}
\startdata
{\rm [Fe/H]} &     0.02 &    0.01 & $-$0.03 \\
{\rm [Al/Fe]} &    0.05 & $-$0.01 & $-$0.01 \\
{\rm [Si/Fe]} & $-$0.01 &    0.03 & $-$0.01 \\
{\rm [Rb/Fe]} &    0.09 &    0.03 & $-$0.01 \\
{\rm [Y/Fe]}  &    0.01 &    0.06 & $-$0.04 \\
{\rm [Zr/Fe]} &    0.13 &    0.02 & $-$0.01 \\
{\rm [La/Fe]} & $-$0.02 &    0.05 & $-$0.05 \\
{\rm [Eu/Fe]} & $-$0.04 &    0.05 & $-$0.04 \\
{\rm [Pb/Fe]} &    0.12 & $-$0.05 & $-$0.02 \\
\enddata

\end{deluxetable}

\end{document}